\documentclass[12pt]{iopart}

\usepackage{amssymb}
\usepackage{graphicx}
\usepackage{dcolumn}
\usepackage{bm}
\usepackage{color}

\newcommand{\prt}{\partial}
\newcommand{\bfr}{{\bf r}}
\newcommand{\om}{\omega}
\newcommand{\vphi}{\varphi}

\begin{document}

\title[Dark soliton oscillations]
{Dark soliton oscillations in Bose-Einstein condensates
with multi-body interactions }

\author{Anatoly M Kamchatnov$^1$ and Mario Salerno$^2$}
\address{
$^1$ Institute of Spectroscopy, Russian Academy of Sciences, Troitsk,
Moscow Region, 142190, Russia \\
$^2$ Dipartimento di Fisica ``E. R. Caianiello'', Consorzio
Nazionale Interuniversitario per le Scienze Fisiche della Materia
(CNISM), Universit\'a di Salerno, via S. Allende I-84081,
Baronissi (SA), Italy }
\ead{kamch@isan.troitsk.ru, salerno@sa.infn.it}

\begin{abstract}
We consider the dynamics of dark matter solitons moving through non-uniform
cigar-shaped Bose-Einstein condensates
described  by the mean field Gross-Pitaevskii equation with
generalized nonlinearities, in the case
when the condition for the modulation stability of the
Bose-Einstein condensate is fulfilled.
The analytical expression for
the frequency of the oscillations of a deep dark soliton is derived
for nonlinearities which are arbitrary functions of the density,
while specific results are  discussed for the physically relevant case
of a  cubic-quintic nonlinearity modeling two- and three-body interactions, respectively.
In contrast to the cubic Gross-Pitaevskii equation for which the frequencies
of the oscillations are known to be independent of background density and
interaction strengths, we find that in the presence of a cubic-quintic nonlinearity
an explicit dependence of the oscillations frequency on the above quantities appears.
This dependence gives rise to the possibility of measuring
these quantities directly from the dark soliton dynamics,
or to manage  the oscillation via the changes of the scattering lengths
by means of Feshbach resonance.
A comparison between analytical results and direct
numerical simulations of the cubic-quintic Gross-Pitaevskii equation
shows good agreement which confirms the validity of our approach.
\end{abstract}


\submitto{\JPB}

\maketitle

\section{Introduction}

Since the  first experimental realization of Bose-Einstein condensates (BECs)
of diluted atomic gases,
the dynamics of matter waves in different trapping potentials
has attracted the attention of many
researchers and presently is a field of growing interest \cite{ps2003, stringari}.
The dispersive properties of BEC induced by the quantum properties
of a slow atomic motion, on one side,
and the  intrinsic nonlinearity of the system induced by the interatomic interactions,
on the other side,
make BECs  ideal systems for  exploring
competing effects of dispersion and nonlinearity. This  results in
a variety of spectacular phenomena ranging from the existences of matter
wave vortices \cite{vortices} and solitons \cite{solitons,Burger1,Densch} to dispersive
shocks \cite{d-shocks,kgk-04}.

In the standard mean field approximation \cite{ps2003} the BEC dynamics is governed
by the Gross-Pitaevskii (GP) equation
\begin{equation}\label{1-1}
    i\hbar\frac{\prt\psi}{\prt t}=-\frac{\hbar^2}{2m}\Delta\psi+V(\bfr)\psi
    +f(|\psi|^2)\psi
\end{equation}
for the condensate ``wave function'' $\psi$, where $V(\bfr)$ is the potential of the
external forces trapping the condensate.  In the standard GP approximation
\begin{equation}\label{1-2}
    f(|\psi|^2)=g|\psi|^2,
\end{equation}
where $g=4\pi\hbar^2a_s/m$
is the nonlinear interaction parameter expressed in terms of the $s$-wave
scattering length $a_s$ of atoms with mass $m$. In this approximation,
only two-body collisions are taken into account. The
scattering length $a_s$, however,  can be varied and even made equal to zero by means
of Feshbach resonances. Then contribution of multi-particle (three-body) collisions
becomes comparable with the two-body contribution and a more complicated
form of the nonlinear term, such as
\begin{equation}\label{2-1}
    f(|\psi|^2)=g|\psi|^2+h|\psi|^4
\end{equation}
arises in the GP equation. Although the three-dimensional GP equation
(\ref{1-1}) is too complicated for analytical studies, one can often consider experimental
low dimensional settings, e.g.  cigar-shaped
traps with a strong radial confinement, for which  the radial motion of the BEC is
``frozen'' and Eq.~(\ref{1-1}) reduces to an effective 1D cubic-quintic generalized
nonlinear Schr\"odinger (NLS) equation
\cite{perez98},
\begin{equation}\label{2-2}
    i\psi_t=-\frac12\psi_{xx}+V(x)\psi+f(|\psi|^2)\psi,
\end{equation}
(here standard non-dimensional variables have been introduced, holding the previous
notation for convenience).
In this reduction, the function $f(\rho),\,\rho=|\psi|^2$,
preserves its form (\ref{2-1}) up to a renormalization of the constants $g$ and $h$.
Also note that the conservation of the norm of the condensate wave function
(normalized number of atoms)
is guaranteed for arbitrary functions $f(\rho)$.
More complicated form of the function $f(\rho)$ can arise if we take into
account the radial motion of BEC (see, e.g., \cite{spr02,ks04}).
Thus, we arrive at the problem
of 1D dynamics of BEC with generalized form of the nonlinearity $f(\rho)$
and confined in the axial direction $x$ by the trap $V(x)$.

As is well known, in case of $f(\rho)$ given by Eq.~(\ref{1-2})
with $g>0$ (repulsive two body interaction) and under the assumption of negligible
changes of the potential $V(x)$ in the region of $x$ under consideration,
the NLS equation
(\ref{2-2}), (\ref{1-2}),  has a dark soliton solution corresponding to the motion
of a hole in the density distribution function, $\rho(x-vt)$,
along a uniform distribution $\rho=\rho_0$ with constant velocity $v$ which
depends on the depth of the soliton \cite{tsuzuki-1971}.
If the potential $V(x)$ is a slowly varying function for distances of the order
of the soliton's width, then
the coordinate of the soliton is well defined and we can speak about
value of the potential $V(x)$ at the ``point'' $x$ of the soliton position.
In this case we can formulate the problem of motion of a dark soliton
``in the potential'' $V(x)$ in terms of its collective coordinate. This problem has
been addressed in several
papers (see, e.g., \cite{ba2000,fpk04,kp04,bkp06,pfk-05,fkp-07,pk-08}) where it was
shown that the dynamics of dark solitons
is quite nontrivial. In particular, if a BEC described by the standard GP equation
(\ref{1-2}) is confined in a harmonic axial trap,
\begin{equation}\label{3-1}
    V(x)=\frac12\om_0x^2,
\end{equation}
then the dark soliton oscillates inside the condensate with the frequency \cite{ba2000}
\begin{equation}\label{3-2}
    \om=\frac{\om_0}{\sqrt{2}}.
\end{equation}
This result is quite surprising because apparently  it contradicts to a simple
consequence of the Ehrenfest theorem which says that the center of mass of the whole
BEC always oscillates with the trap frequency $\om_0$. The remarkable theoretical
prediction (\ref{3-2}) was confirmed in the experiment \cite{becker}.
Equation (\ref{3-2}) then implies
that the motion of the dark soliton must be accompanied by a deformation
of the density distribution which re-tunes the oscillation of the whole condensate to
the trap frequency. In this situation the motion of the dark soliton  becomes
``decoupled'' from the motion of the center of mass and the Ehrenfest
theorem is not violated.

However, the result (\ref{3-2}) is not universal.  If the nonlinearity
function is given by
\begin{equation}\label{3-3}
f(\rho)=h\rho^2
\end{equation}
with $h>0$, then the frequency $\om$ of oscillations of a shallow dark soliton
coincides with the trap frequency \cite{fpk04},
\begin{equation}\label{3-4}
    \om=\om_0
\end{equation}
and for deep solitons this frequency is given by a more complicated expression
(see \cite{bkp06})
\begin{equation}\label{3-5}
    \om=\left\{1+\frac{\sqrt{3}}{2\ln\frac{1+\sqrt{3}}{\sqrt{2}}}\right\}^{-1/2}\om_0
    \approx 0.6572 \om_0.
\end{equation}
This observation poses the problem of the dark soliton motion in the case of
a more general form of the nonlinearity.

The aim of the  present paper is to address this  problem by considering a non-uniform
cigar-shaped BEC with nonlinear interactions modeled by an arbitrary  function $f(\rho)$
of the density. In particular we consider the physically relevant example of the
cubic-quintic  nonlinearity function in  (\ref{2-1}),  modeling the
mean field effect of the two-body and three-body collisions in the GP equation.
In this case  we find that the relative
contribution of the two nonlinear terms to the soliton oscillation depends on the
background density $\rho_0$:
if a deep soliton oscillates in vicinity of the maximum $\rho_0$ of the
density distribution, then for $|h|\rho_0\ll g$ its frequency must be
equal to (\ref{3-2}), and for $|h|\rho_0\gg g$ to (\ref{3-5}). The
transition from one limiting case to another is of considerable
theoretical interest.
We derive an  analytical expression for
the oscillation frequency of a deep dark solitons
which reveals  the explicit
dependence of the oscillations frequency both on the
background density and on the strength of the nonlinearity parameters.
Our formulae reproduce the results   considered in \cite{kp04,bkp06} for the
particular cases (\ref{1-2}) and (\ref{3-3})
of the nonlinearity for which  the above  dependence disappears.

We also consider the case of a quintic attractive nonlinearity ($h<0$) with
an overall repulsive interaction.
In this case for the density $\rho>g/(2|h|)$ we get modulation
instability of BEC, i.e. the sound velocity
\begin{equation}\label{3-6}
    c_s=\sqrt{\rho_0f'(\rho_0)}
\end{equation}
becomes imaginary so that the
notion of ``dark solitons'' looses its sense.  We show that in this case the oscillation
frequency $\om$ as a function of $\rho_0$ becomes singular
at $\rho_{0c}=g/(2|h|)$. The form of this singularity may provide further advances
in the  understanding of the dynamics of dark solitons in BEC.

Our theory will be  based on the elegant approach of Refs.~\cite{kp04,bkp06}
where the dark soliton was considered as a ``Landau quasi-particle''
whose motion in the stationary potential $V(x)$ is governed by the
quasi-particle's ``energy conservation'' law in close analogy with
the Newtonian particle motion.
The generalization of this approach
to arbitrary $f(\rho)$ allows us to derive the equation of motion of the dark soliton
for the general case and to show the dependence of the soliton effective  mass
on the system parameters.
A comparison between analytical results and direct
numerical simulations of the cubic-quintic GP equation shows a very good agrement
which further confirms the validity of our
analysis.

We finally remark that the dependence of the dark  soliton oscillation frequency
on the system parameters
can be very useful for experimental studies
of relative contribution of multi-particle collisions into BEC
dynamics by varying the value of $g$ by means of Feshbach resonance.
Moreover, it gives the possibility  to  measure background densities or scattering
lengths directly from the dark soliton dynamics. The possibility to  manage soliton
oscillations via temporal
changes of the scattering lengths is also an attractive  feature emerging from
our study which could be
of interest for practical applications.

The paper is organized as follows. In Section II we discuss the general theory
of the dynamics
of a dark soliton of the  NLS equation with an arbitrary nonlinearity. The results
will be used to study the cases of both
deep and shallow soliton dynamics. In Section III we consider oscillations
of dark solitons for the specific case of
a cubic-quintic  nonlinearity. We consider  the cases  of a repulsive cubic
nonlinearity and either repulsive or  attractive
quintic nonlinearity, the last  being done for parameters satisfying the
condition of modulation stability of BEC.
In Section IV analytical results will be compared with direct numerical
simulations of the cubic-quintic GP equation for the case  of
repulsive and attractive cubic and quintic nonlinearity, respectively.
In the last Section we draw our conclusions and summarize
the main results of the paper.

\section{Motion of a dark soliton in a trap: general theory}

First, we have to find the soliton solution of the generalized NLS
equation (\ref{2-2}) with zero potential $V(x)\equiv0$ but under
the boundary condition that the condensate density $\rho(x)$ tends to
the limiting background density $\rho_0$ as $|x|\to\infty$ and the
condensate's velocity vanishes at infinity. To this end, it is
convenient to look for the solution in the form
\begin{equation}\label{4-1}
    \psi=\sqrt{\rho(\xi)}\exp\left(i\vphi(\xi)-i\mu t\right)
\end{equation}
where
\begin{equation}\label{4-2}
    \mu=f(\rho_0)
\end{equation}
is the chemical potential of the condensate and its density $\rho$
and the local flow velocity $u=\vphi_x$ depend only on the variable
$\xi=x-vt$ with $v$ being the soliton velocity. The formulated
above boundary conditions take the form
\begin{equation}\label{4-3}
    \rho\to\rho_0,\quad \vphi_{\xi}\to0,\quad\mathrm{as}\quad
    |\xi|\to \infty.
\end{equation}
Substitution of (\ref{4-1}) into (\ref{2-2}) with $V(x)\equiv0$ and separation of
real and imaginary parts yields the system of two equations for
$\rho(\xi)$ and $u(\xi)=\vphi_\xi$. One of these equations can be
integrated at once to give
\begin{equation}\label{4-4}
    u(\xi)=\vphi_\xi=v\left(1-\frac{\rho_0}{\rho}\right)
\end{equation}
where the integration constant is chosen according to the conditions
(\ref{4-3}). Then the second equation can be transformed to
\begin{equation}\label{5-1}
    \frac18\rho_{\xi}^2-\frac14\rho\rho_{\xi\xi}+\rho^2f(\rho)-\mu\rho^2
    +\frac12v^2(\rho_0^2-\rho^2)=0
\end{equation}
and this equation can also be integrated once. As a result we obtain
the equation
\begin{equation}\label{5-2}
    \rho_{\xi}^2=Q(\rho)
\end{equation}
where
\begin{equation}\label{5-3}
    Q(\rho)=8\rho\int_{\rho}^{\rho_0}[\mu-f(\rho')]d\rho'-4v^2(\rho_0-\rho)^2
\end{equation}
and again the integration constant is chosen according to the condition
(\ref{4-3}). Thus, we have reduced the problem of finding the density
profile to the inversion of the integral
\begin{equation}\label{5-4}
    \xi=\int_{\rho_m}^\rho\frac{d\rho}{\sqrt{Q(\rho)}}
\end{equation}
where $\rho_m$ is the minimal density at the center $\xi=0$ of the soliton.
The function $Q(\rho)$ has a double zero at $\rho=\rho_0$ and the condition
$\left.dQ(\rho)/d\rho\right|_{\rho=\rho_0}=0$ reproduces the relation (\ref{4-2}).
At $\rho=\rho_m$ the function $Q(\rho)$ has a single zero, and this condition
yields the relationship between $\rho_m$ and the soliton velocity $v$,
\begin{equation}\label{5-5}
    v^2=\frac{Q_0(\rho_m)}{4(\rho_0-\rho_m)^2},
\end{equation}
where
\begin{equation}\label{5-5a}
    Q_0(\rho)=8\rho\int_{\rho}^{\rho_0}[f(\rho_0)-f(\rho')]d\rho'
\end{equation}
is the function (\ref{5-3}) with zero value of the velocity $v$
and we have used here Eq.~(\ref{4-2}) for $\mu$.

Now we have to obtain the expression for the energy of the dark soliton.
To this end, we notice that the function $\Psi=\sqrt{\rho}\exp(i\vphi)=
\psi\cdot\exp(i\mu t)$ satisfies the equation
\begin{equation}\label{5-6}
    i\Psi_t+\frac12\Psi_{xx}+(\mu-f(|\Psi|^2))\Psi=0
\end{equation}
which can be written in a Hamiltonian form
\begin{equation}\label{5-7}
    i\Psi_t=\frac{\delta H}{\delta\Psi^*}
\end{equation}
with
\begin{equation}\label{5-8}
    H=\int_{-\infty}^{\infty}\left[\frac12|\Psi_x|^2+\int_0^{|\Psi|^2}
    f(\rho)d\rho-\mu|\Psi|^2\right]dx.
\end{equation}
To get the contribution of the dark soliton into the energy of BEC, we subtract
the background energy $H_0=\int_{-\infty}^{\infty}\left[\int_0^{\rho_0}
f(\rho)d\rho-\mu|\Psi|^2\right]dx$
and express the resulting formula in terms of $\rho$ and $\vphi_{x}$,
\begin{equation}\label{6-1}
    E=H-H_0=\int_{-\infty}^{\infty}\left[\frac12\left(\vphi_x^2+\frac{\rho_x^2}{4\rho^2}
    \right)\rho+\int_{\rho}^{\rho_0}
    [f(\rho_0)-f(\rho')]d\rho'\right]dx.
\end{equation}
Substitution of expressions (\ref{4-4}) and (\ref{5-2}) for the dark soliton
solution casts this formula into
\begin{equation}\label{6-2}
    E=2\int_{-\infty}^{\infty} \int_{\rho}^{\rho_0}[f(\rho_0)-f(\rho')]d\rho'dx.
\end{equation}
At last, integration with respect to $x$ can be transformed to integration with respect to
$\rho$ from $\rho_m$ to $\rho_0$ with account of (\ref{5-4}),
\begin{equation}\label{6-3}
    E=4\int_{\rho_m}^{\rho_0}\frac{d\rho}{\sqrt{Q(\rho)}}\int_{\rho}^{\rho_0}
    [f(\rho_0)-f(\rho')]d\rho'
    =\frac12\int_{\rho_m}^{\rho_0}\frac{Q_0(\rho)}{\rho\sqrt{ Q(\rho)}}d\rho.
\end{equation}
Here $\rho_m$ can be considered according to (\ref{5-5}) as a function of $\rho_0$
and $v^2$, hence the soliton energy is also a function of these two variables,
\begin{equation}\label{6-4}
E=E(\rho_0,v^2).
\end{equation}

Now we turn to the problem of motion of the dark soliton along a nonuniform condensate
confined in a trap with the axial potential $V(x)$. Then in the Thomas-Fermi (TF)
approximation the density $\rho(x)$ is a function of $x$ determined implicitly by the equation
\begin{equation}\label{6-5}
    \mu -V(x)=f(\rho(x)).
\end{equation}
When this TF distribution of the density $\rho=\rho(x)$ is substituted into
(\ref{6-4}) instead of $\rho_0$, we obtain the energy of the soliton moving
with velocity $v$ and located at this moment at the point $x$. As was shown in
Ref.~\cite{kp04}, the velocity $v$ changes during the motion of the dark soliton
in such a way that the soliton energy (\ref{6-4}) is conserved. If we denote
the soliton coordinate as $x=X(t)$, then $v=\dot{X}(t)$ and Eq.~(\ref{6-4})
converts into
\begin{equation}\label{6-6}
    E(\rho(X),\dot{X}^2)=E_0,
\end{equation}
where $E_0$ is the initial energy of the soliton. Since the function $E(\rho,v^2)$ is
known, Eq.~(\ref{6-6}) is the differential equation for finding the soliton
coordinate $X$ as a function of time $t$. In general, this first order
ordinary differential equation should be solved numerically. Here we shall
consider the most interesting limiting cases admitting complete analytical study.

\subsection{Small amplitude oscillations of a deep soliton}

Let us consider oscillations of a deep soliton,
\begin{equation}\label{7-1}
    \rho_m\ll\rho_0,
\end{equation}
where $\rho_0$ is the maximal density of the condensate at its ``top'' when
it is confined in a trap with the potential $V(x)$ (see (\ref{6-5})).
In this case the amplitude of oscillations is much less than the TF radius
of the condensate and the velocity of its motion is much less than the
local sound velocity
\begin{equation}\label{7-4}
    v\ll c_s,\quad c_s=\sqrt{\rho_0f'(\rho_0)}.
\end{equation}
Therefore, if we expand the energy into series with respect to small amplitude
and velocity, then the first nontrivial terms in (\ref{6-6}) will give
the expression
\begin{equation}\label{7-5}
    \frac12 m_*\dot{X}^2+V(X)=\mathrm{const},
\end{equation}
where $V(X)$ changes little for the small amplitude vibrations. Hence, we
get the ``energy conservation law'' for a Newtonian particle moving in
the potential $V(X)$, where $m_*$ is the effective mass of the dark soliton.

For small $v$ we obtain from (\ref{5-5})
\begin{equation}\label{7-6}
    \rho_m\cong\frac{\rho_0^2v^2}{2a}
\end{equation}
where
\begin{equation}\label{7-7}
    a=\int_0^{\rho_0}[f(\rho_0)-f(\rho)]d\rho.
\end{equation}
To estimate the first term of the series expansion of $E$ with respect to $v^2$,
we split the integral (\ref{6-3}) into two terms (see \cite{bkp06}),
\begin{equation}\label{8-1}
    \begin{array}{l}
    E=E_1+E_2,\\
    E_1=\frac12\int_{\rho_m}^{\rho_1}\frac{Q_0(\rho)}{\rho\sqrt{ Q(\rho)}}d\rho,\quad
    E_2=\frac12\int_{\rho_1}^{\rho_0}\frac{Q_0(\rho)}{\rho\sqrt{ Q(\rho)}}d\rho,
    \end{array}
\end{equation}
where we have introduced an intermediate integration limit $\rho_1$ so that
\begin{equation}\label{8-2}
    \rho_m\ll\rho_1\ll\rho_0.
\end{equation}
Dependence of $E_1$ on $v^2$ results mainly from the lower integration limit
(\ref{7-6}) and in the main approximation we get
\begin{equation}\label{8-3}
    E_1\cong\frac12\int_{\rho_m}^{\rho_1}\sqrt{Q(\rho)}\frac{d\rho}{\rho}.
\end{equation}
Now, both limits of integration are small and $\rho_m$ is a simple zero
of the function $Q(\rho)$, hence we can approximate it as
\begin{equation}\label{8-4}
    Q(\rho)\cong\left.\frac{dQ(\rho)}{d\rho}\right|_0(\rho-\rho_m)=8a(\rho-\rho_m),
\end{equation}
where $a$ is defined in (\ref{7-7}). Then substitution of (\ref{8-4}) into
(\ref{8-3}) and elementary integration yield
\begin{equation}\label{8-5}
    E_1\cong\sqrt{8a(\rho_1-\rho_m)}.
\end{equation}
We need the coefficient of $v^2$ in the expansion of $E$ in powers of $v^2$
and the contribution of $E_1$ into this coefficient is equal to
\begin{equation}\label{8-6}
    \frac{dE_1}{d\rho_m}\frac{d\rho_m}{dv^2}=-\frac{\rho_0^2}{\sqrt{2a\rho_1}}.
\end{equation}
As we see, it diverges in the limit $\rho_1\to0$. To cancel this divergent term,
we have to identify a similar term in the derivative $dE_2/dv^2$ given by the expression
\begin{equation}\label{9-1}
    \left.\frac{dE_2}{dv^2}\right|_{v=0}=\int_{\rho_1}^{\rho_0}
    \frac{(\rho_0-\rho)^2}{\rho\sqrt{Q_0(\rho)}}d\rho.
\end{equation}
According to Eq.~(\ref{8-6}), the diverging term is proportional to $\rho_1^{-1/2}$.
Correspondingly, we transform the integral in (\ref{9-1}) in the following way
by integration by parts,
\begin{equation}\label{9-3}
\begin{array}{ll}
    \left.\frac{dE_2}{dv^2}\right|_{v=0}&=\int_{\rho_1}^{\rho_0}
    \frac{(\rho_0-\rho)^2}{\rho^{3/2}\sqrt{Q_0(\rho)/\rho}}d\rho=
    \int_{\rho_1}^{\rho_0}
    \frac{(\rho_0-\rho)^2}{\sqrt{Q_0(\rho)/\rho}}d(-2\rho^{-1/2})\\
    &= -\left.\frac{2(\rho_0-\rho)^2}{\sqrt{Q_0(\rho)}}\right|_{\rho_1}^{\rho_0}+
    \int_{\rho_1}^{\rho_0}\frac2{\sqrt{\rho}}\frac{d}{d\rho}\left[
    \frac{\sqrt{\rho}(\rho_0-\rho)^2}{\sqrt{Q_0(\rho)}}\right]d\rho.
    \end{array}
\end{equation}
The first term vanishes at the upper limit $\rho=\rho_0$ and at the lower limit
$\rho=\rho_1\ll\rho_0$ we can approximate $Q_0(\rho)$ as $Q_0(\rho)\cong8a\rho$
(see (\ref{8-4})). In the second term the integral is convergent and we can take
here the lower integration limit equal to $\rho_1=0$. As a result we get
\begin{equation}\label{9-4}
    \left.\frac{dE_2}{dv^2}\right|_{v=0}\cong\frac{\rho_0^2}{\sqrt{2a\rho_1}}+
    2\int_{0}^{\rho_0}\frac{d\rho}{\sqrt{\rho}}\frac{d}{d\rho}\left[
    \frac{\sqrt{\rho}(\rho_0-\rho)^2}{\sqrt{Q_0(\rho)}}\right].
\end{equation}
The first term cancels with the contribution (\ref{8-6}) and we obtain the
expression for the soliton mass
\begin{equation}\label{9-5}
    m_s=2\left.\frac{dE_2}{dv^2}\right|_{v=0}=
    4\int_{0}^{\rho_0}\frac{d\rho}{\sqrt{\rho}}\frac{d}{d\rho}\left[
    \frac{\sqrt{\rho}(\rho_0-\rho)^2}{\sqrt{Q_0(\rho)}}\right]
\end{equation}
for the motion of a dark soliton near the ``top'' of the condensate.

Introduction of the potential $V(x)$ leads to the shift of the chemical potential
$\mu\to\mu-V(x)$. Taking into account the thermodynamic formula
\begin{equation}\label{9-6}
    N=-\frac{dE}{d\mu}
\end{equation}
and the expression for the energy in the limit $v^2\to0$,
\begin{equation}\label{9-7}
\begin{array}{ll}
   E=E_0&=4\int_0^{\rho_0}\frac{d\rho}{\sqrt{Q_0(\rho)}}\int_{\rho}^{\rho_0}
    [\mu-f(\rho')]d\rho'\nonumber  \\
    &= 4 \int_0^{\rho_0}\frac{d\rho}{\sqrt{8\rho}}\left\{\int_{\rho}^{\rho_0}
    [\mu-f(\rho')]d\rho'\right\}^{1/2},
    \end{array}
\end{equation}
hence
\begin{equation}\label{10-1}
    N=-2\int_0^{\rho_0}\frac{(\rho_0-\rho)d\rho}{\sqrt{Q_0(\rho)}},
\end{equation}
we find the expansion of the soliton energy as
\begin{equation}\label{10-2}
    E=E_0+\left.\frac{dE}{dv^2}\right|_{v^2=0}\cdot v^2+\frac{dE}{d\mu}\cdot(-V(x))=
    E_0+\frac12m_sv^2+NV(x),
\end{equation}
and division by $N$ yields the energy conservation law for the soliton's motion
\begin{equation}\label{10-3}
    \frac{m_*}2v^2+V(x)=\mathrm{const}
\end{equation}
where
\begin{equation}\label{10-4}
    m_*=\frac{m_s}N
\end{equation}
is the effective mass of the soliton's motion.
Hence, in case of a harmonic trap (\ref{3-1}) we get the formula
\begin{equation}\label{10-5}
    \om=\frac{\om_0}{\sqrt{m_*}}
\end{equation}
for the frequency of oscillations of deep solitons.

For a usual GP equation we have
\begin{equation}\label{10-6}
    f(\rho)=\rho,
\end{equation}
hence
\begin{equation}\label{10-7}
    Q_0(\rho)=4\rho(\rho_0-\rho)^2,\quad m_s=-4\sqrt{\rho_0},\quad
    N=-2\sqrt{\rho_0}
\end{equation}
and
\begin{equation}\label{10-8}
    m_*=2
\end{equation}
so that (\ref{10-5}) reproduces the well-known result.

\subsection{Motion of a shallow soliton}

In the shallow soliton limit
\begin{equation}\label{10-9}
    \rho_0-\rho\ll\rho_0
\end{equation}
the series expansion of Eq.~(\ref{5-5}) gives
\begin{equation}\label{10-10}
    v^2\cong\rho_m\left[f'(\rho_0)-
    \frac13f^{\prime\prime}(\rho_0)(\rho_0-\rho_m)\right]
\end{equation}
or with account of the expression (\ref{3-6}) for the sound velocity
we obtain
\begin{equation}\label{10-11}
    \rho_m=\rho_0-\frac{c_s^2-v^2}{f'(\rho_0)+
    \frac13\rho_0f^{\prime\prime}(\rho_0)}.
\end{equation}

In this limit the dark soliton can be considered in the Korteweg-de Vries (KdV)
approximation (see, e.g. \cite{kl-d98}) for which the NLS equation with
$V(x)\equiv0$ reduces to the equation
\begin{equation}\label{11-1}
    \rho'_t+c_s\rho'_x+\frac3{2c_s}\left[f'(\rho_0)+
    \frac13\rho_0f^{\prime\prime}(\rho_0)\right]\rho'\rho'_x-
    \frac1{8c_s}\rho'_{xxx}=0
\end{equation}
for evolution of small deviations $\rho'=\rho_0-\rho$ of the density
from the background value $\rho_0$. This equation has the soliton
solution
\begin{equation}\label{11-2}
    \rho'=\rho_0-\rho=\frac{c_s^2-v^2}{f'(\rho_0)+
    \frac13\rho_0f^{\prime\prime}(\rho_0)}\cdot\frac1
    {\cosh^2[\kappa(x-vt)]}
\end{equation}
where
\begin{equation}\label{11-3}
    \kappa=2\sqrt{\frac{c_s(c_s^2-v^2)}{f'(\rho_0)+
    \frac13\rho_0f^{\prime\prime}(\rho_0)}}
\end{equation}
is the inverse soliton's half-width and the minimal density $\rho_m$ at the
soliton's center $x=vt$ is given by the expression (\ref{10-11}).

In the same approximation
\begin{equation}\label{11-4}
    Q(\rho)=4\left[f'(\rho_0)+\frac13\rho_0f^{\prime\prime}(\rho_0)\right]
    (\rho_0-\rho)^2(\rho-\rho_m)
\end{equation}
and calculation of the soliton energy (\ref{6-3}) gives
\begin{equation}\label{11-5}
    E=\frac43\frac{f'(\rho_0)}{\left[f'(\rho_0)+
    \frac13\rho_0f^{\prime\prime}(\rho_0)\right]}\cdot(c_s^2-v^2)^{3/2}.
\end{equation}

If the width $1/\kappa$ of the soliton is much less than the condensate's
size (the TF radius), then we can consider the motion of such a soliton in
the trap by replacement of $\rho_0$ by the solution $\rho(x)$ of
Eq.~(\ref{6-5}) and of $v$ by its variable velocity $dX/dt$ to get the equation
\begin{equation}\label{11-6}
    \left(\frac{dX}{dt}\right)^2=\rho f'(\rho)-\left\{\frac{3E}{4f'(\rho)}
    \left[f'(\rho)+\frac13f^{\prime\prime}(\rho)\right]^2\right\}^{2/3},
\end{equation}
where $\rho=\rho(X)$ is given by Eq.~(\ref{6-5}). For example, in case of
$f=h\rho^2,\,h>0,$ we have $f'(\rho)=2h\rho$ and $\mu-V(X)=h\rho^2$, hence (\ref{11-6})
reduces to
\begin{equation}\label{11-7}
    \frac12\left(\frac{dX}{dt}\right)^2+V(X)+2\left(\frac{hE}3\right)^{2/3}
    \left(\frac{\mu-V(X)}h\right)^{1/3}=\mu.
\end{equation}
As we see, for a harmonic trap (\ref{3-1}) the frequency of oscillations depends
on the energy $E$ of the soliton. However, if $E$ is small enough,
\begin{equation}\label{12-1}
    E\ll\frac{\mu}{\sqrt{h}},
\end{equation}
then the last term in the left-hand size can be neglected and we return to the
usual energy conservation law for a Newtonian particle which for a harmonic trap
gives Eq.~(\ref{3-4}) in agreement with \cite{fpk04,bkp06}.

\section{Oscillation of a dark soliton for the cubic-quintic nonlinearity}

Now we shall consider in some detail the motion of a dark soliton described by
the generalized NLS equation with $f(\rho)$ given by Eq.~(\ref{2-1}), i.e.,
\begin{equation}\label{12-2}
    f(\rho)=g\rho+h\rho^2.
\end{equation}
Conditions for realization of motion of a shallow soliton are very restrictive and
we shall confine ourselves to the most practical case of small oscillations of
a deep soliton (Sec. 2A).

Substitution of (\ref{12-2}) into (\ref{5-5a}) gives
\begin{equation}\label{12-3}
    Q_0(\rho)=8\rho(\rho_0-\rho)^2\left[\frac12g+\frac13h(2\rho_0+\rho)\right].
\end{equation}
Situations with $h>0$ and $h<0$ should be considered separately.

\begin{figure}[bt]
\centerline{
\includegraphics[width=7cm,clip]{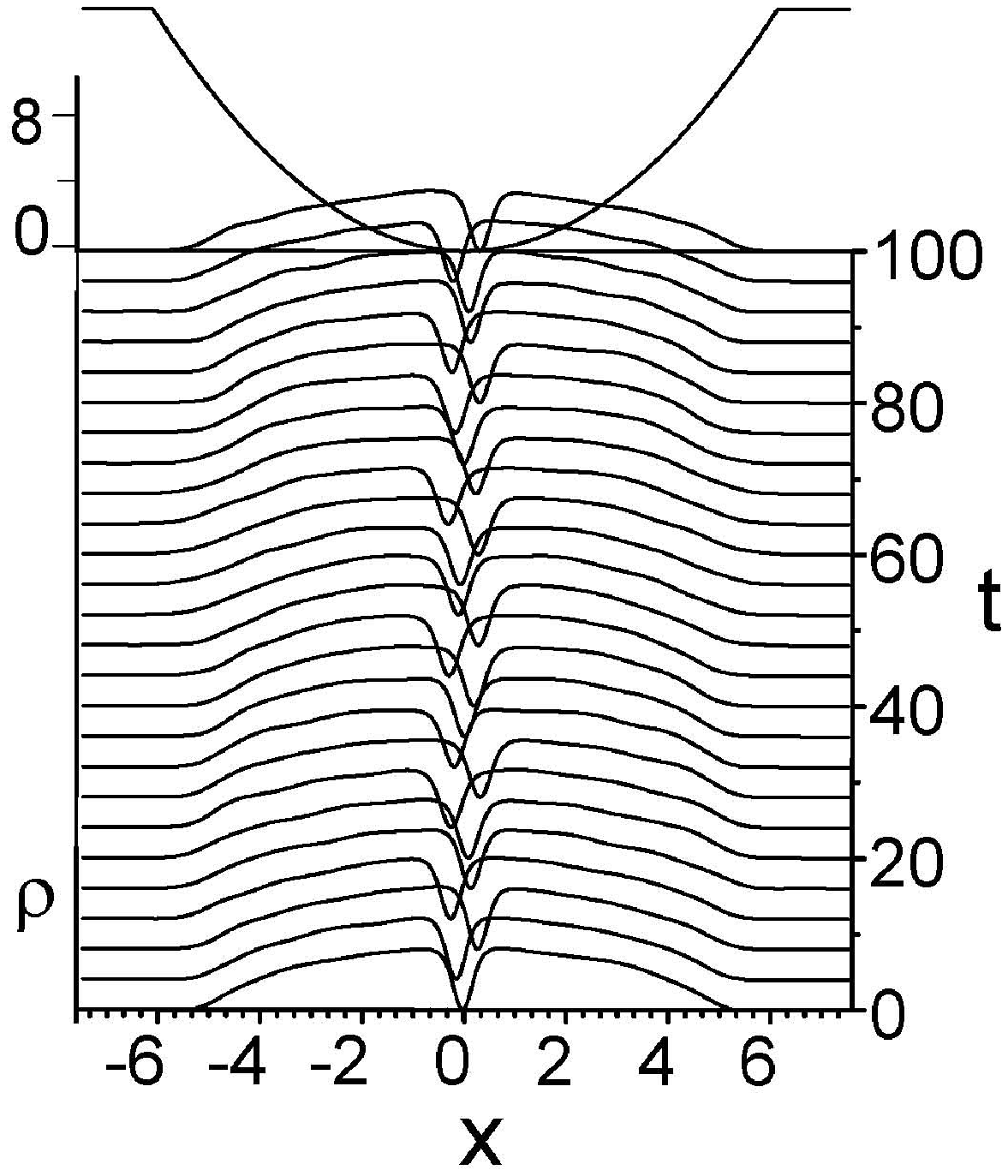}\qquad
\includegraphics[width=7cm,clip]{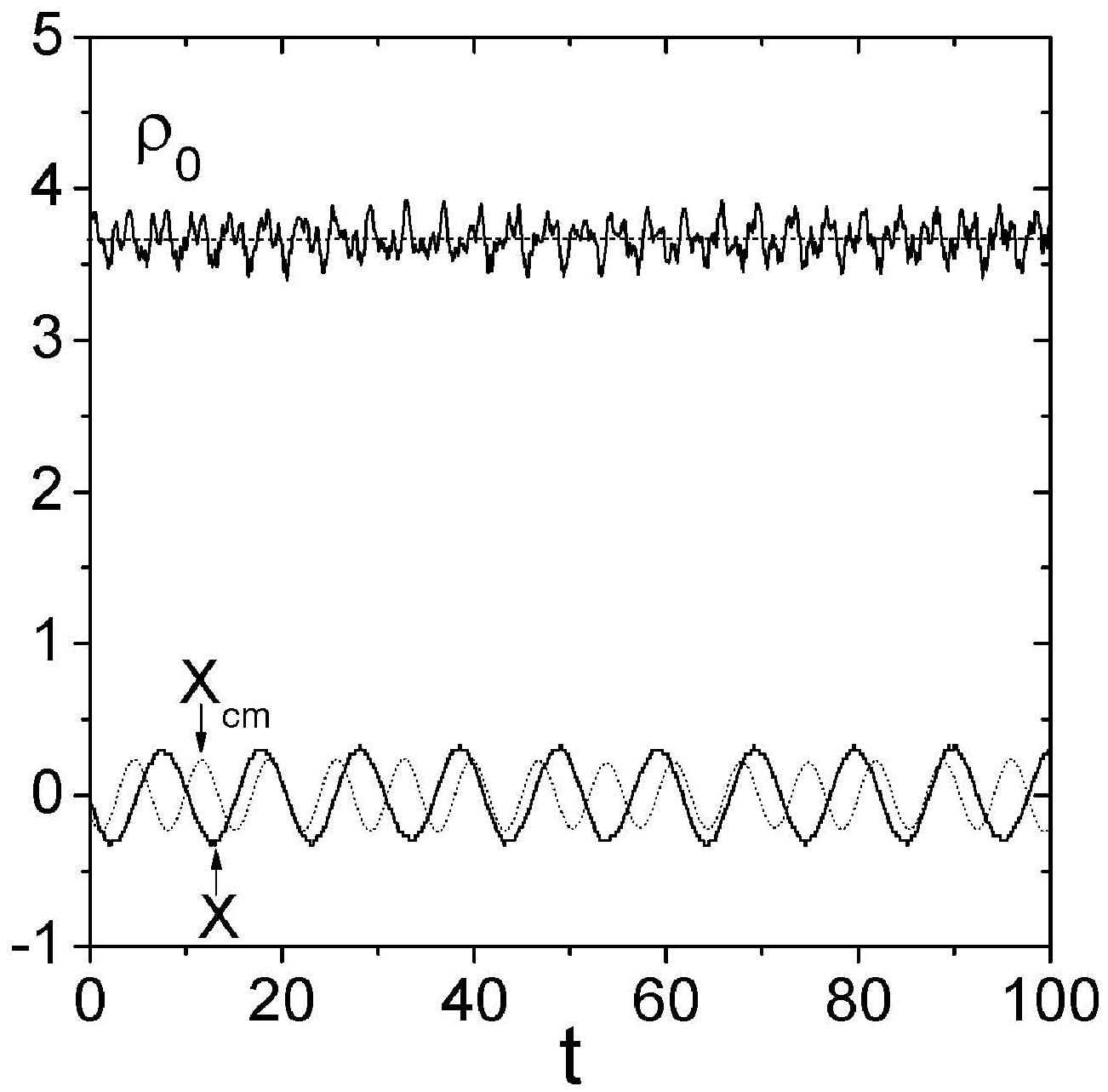}
}
\caption{Left panel. Time evolution of a deep dark soliton in a BEC with repulsive
two- and three-body interactions obtained from numerical simulation of the
cubic-quintic GPE with
$g=1.0, h=0.5$ and parabolic potential $V(x)=\frac 12 \omega_0^2$
with $\omega_0=0.894$ (the potential is depicted at the top of the figure
for scale comparison).
Right panel. Dark soliton center coordinate (bottom curve) and soliton
density background corresponding to
the  dynamics depicted in the right panel. The numerical oscillation frequency
of the soliton is $\omega=0.607$, while the time averaged background density,
indicated by
the dashed line in the figure,
is $\langle\rho_0\rangle=3.674$. The thin dotted line at the bottom refers
to the dynamics of the
center of mass of the whole condensate.}
\label{fig1}
\end{figure}

\subsection{Repulsive quintic interaction ($h>0$)}

In this case elementary integration of Eq.~(\ref{5-4}) with $Q(\rho)$ given
by (\ref{12-3}) yields the profile of the density in the dark solitons solution
\begin{equation}\label{sol1}
    \rho(\xi)=\rho_0-\frac{(\rho_0-\rho_+)(\rho_0-\rho_-)}
    {(\rho_+-\rho_-)\cosh^2(\kappa\xi)+\rho_0-\rho_+},\quad \xi=x-vt,
\end{equation}
where
\begin{equation}\label{sol2}
    \rho_{\pm}=\pm\sqrt{\left(\frac{3g}{4h}+\rho_0\right)^2+\frac{3v^2}{2h}}
    -\left(\frac{3g}{4h}+\rho_0\right)
\end{equation}
and
\begin{equation}\label{sol3}
    \kappa=\sqrt{2h(\rho_0-\rho_+)(\rho_0-\rho_-)/3}.
\end{equation}
Substitution of (\ref{12-3}) into (\ref{10-1}) and
simple calculation give
\begin{equation}\label{12-4}
    N=-\sqrt{\frac6h}\ln\left(\sqrt{\frac{2h\rho_0}{3g+4h\rho_0}}+
    \sqrt{\frac{3(g+2h\rho_0)}{3g+4h\rho_0}}
    \right).
\end{equation}
Similar calculation of (\ref{9-5}) yields
\begin{equation}\label{12-5}
    m_s=-\sqrt{\frac6h}\left[\ln\left(\sqrt{\frac{2h\rho_0}{3g+4h\rho_0}}+
    \sqrt{\frac{3(g+2h\rho_0)}{3g+4h\rho_0}}\right)
    +\frac{\sqrt{6h\rho_0(g+2h\rho_0)}}{3g+4h\rho_0}\right].
\end{equation}
Hence the effective mass of the soliton motion is equal to
\begin{equation}\label{12-6}
    m_*=1+\frac{\sqrt{6h\rho_0(g+2h\rho_0)}}{(3g+4h\rho_0)
    \ln\left(\sqrt{\frac{2h\rho_0}{3g+4h\rho_0}}+
    \sqrt{\frac{3(g+2h\rho_0)}{3g+4h\rho_0}}
    \right)}.
\end{equation}
For $h\rho_0\ll g$ this formula reproduces
\begin{equation}\label{13-1}
    m_*=2,\quad h\rho_0\ll g,
\end{equation}
in agreement with (\ref{3-2}), and for $h\rho_0\gg g$ we get
\begin{equation}\label{13-2}
    m_*=1+\frac{\sqrt{3}}{2\ln\frac{1+\sqrt{3}}{\sqrt{2}}}, \quad h\rho_0\gg g,
\end{equation}
in agreement with (\ref{3-5}).

\begin{figure}[bt]
\centerline{
\includegraphics[width=7cm,height=7cm,clip]{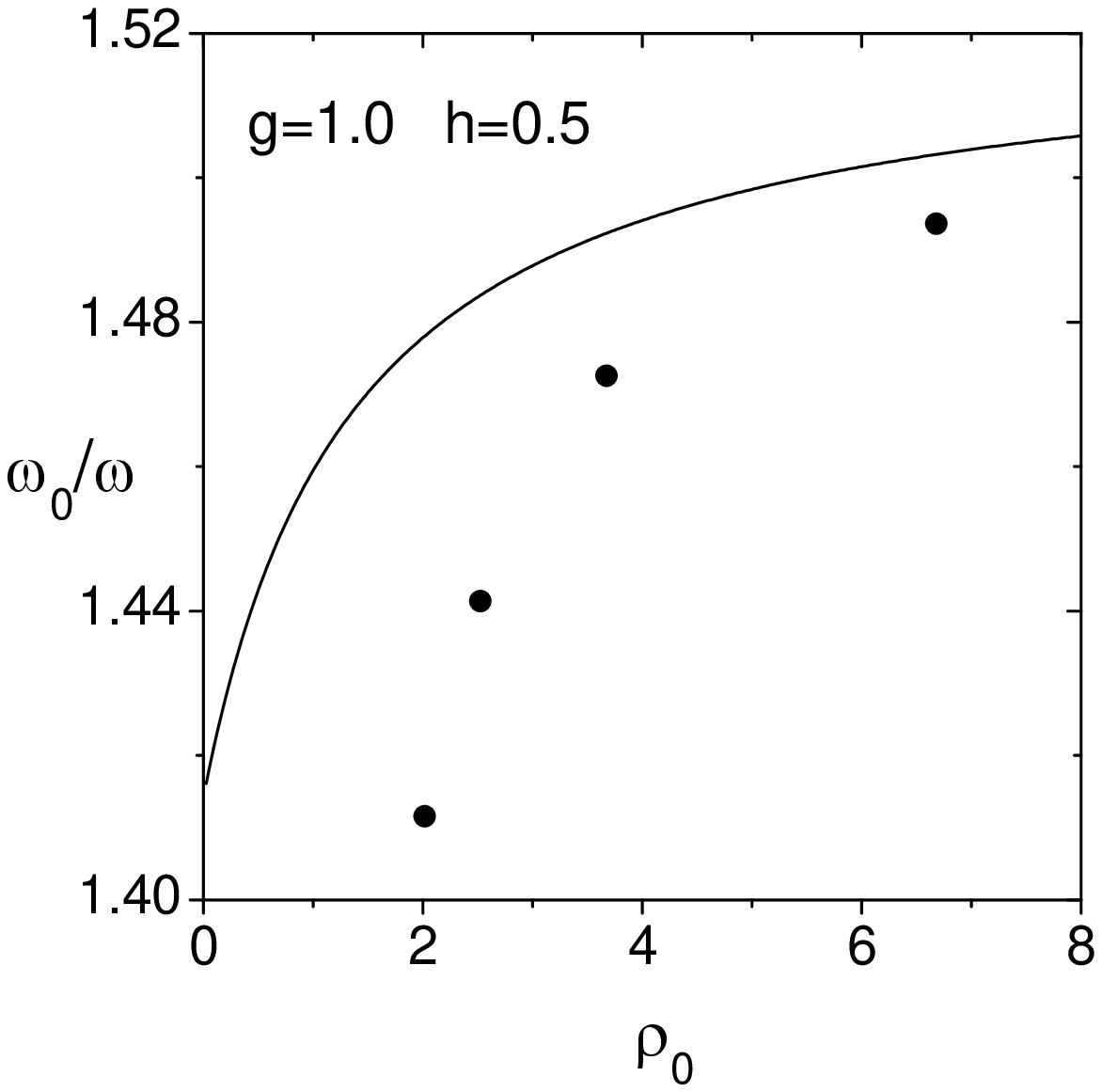}
\includegraphics[width=7cm,height=7cm,clip]{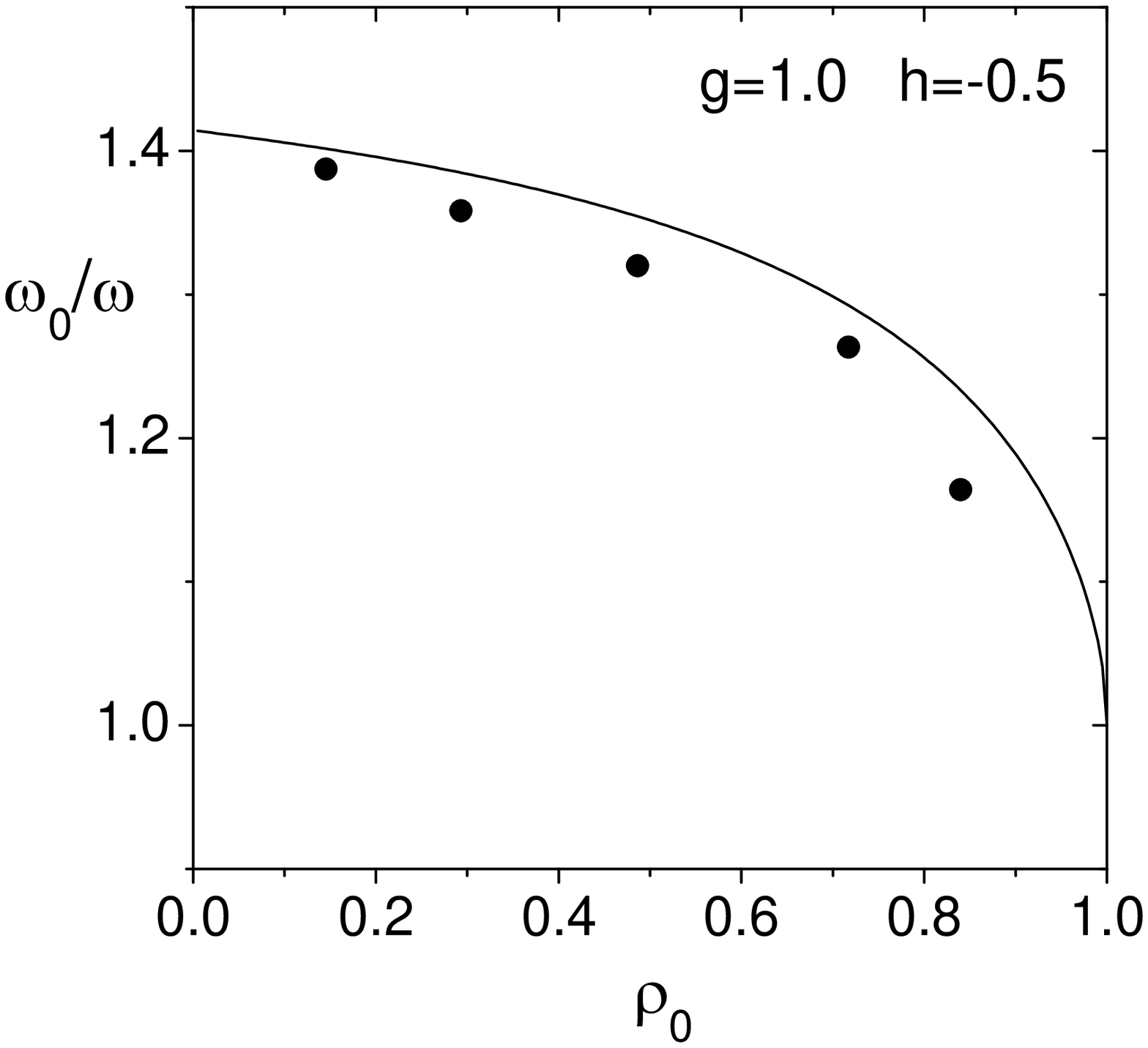}
}
\caption{Dependence of the ratio $\om_0/\om=\sqrt{m_*}$ on the BEC background
density $\rho_0$ for repulsive  interactions, $g=1.0,\,h=0.5$, (left panel) and
for opposite sign interactions $g=1.0,\,h= - 0.5$ (right panel).
Continuous curves refer to the analytical results
in Eqs.~(\ref{12-6}), (\ref{13-6}), while the dots are obtained from
direct numerical solution of
the generalized NLS equation (\ref{2-2})
with $f(\rho)$ given by Eq.~(\ref{12-2}).}
\label{fig2}
\end{figure}

To check these results, we have numerically integrated the GP equation for this
case using as
initial condition the exact dark soliton solution of the cubic-quintic NLS in
Eq.~(\ref{1-1}).
The parabolic potential was introduced adiabatically by rising with time
the frequency $\omega_0$
from zero (pure cubic-quintic solution) up to a desired value. Effects of adiabatic
changes of parameters on BEC
bright and dark solitons were considered in  Ref.~\cite{AS03} and the
effectiveness of the method
was demonstrated. In Fig.~\ref{fig1} we depict the dynamics of a repulsive
condensate with its dark soliton inside
in presence of a  parabolic trap raised adiabatically as described above.
In the right panel of this
figure we show the dynamics of the center position $X(t)$ of the dark soliton and of the
center of mass of the whole condensate $X_{cm} = \int x |\psi|^2 dx$ together with
the time evolution of the background density at the soliton position.  We see that
both the position of the soliton and of the BEC center of mass are very
regular and relatively small amplitude  excitations develop on the background during
the evolution.
Note that  the center of mass  of the condensate oscillates with the trap frequency
in agreement with Ehrenfest theorem, while
the oscillation frequency and the effective mass of the dark soliton are in very
good  agreement
with our analytical predictions $\omega=0.599, m^*=2.227$ (for the comparison with
numerics we take as $\rho_0$ the
time average of the soliton background
measured during the evolution).
A more extended comparison between the theory and
numerical results is presented in the left panel
of Fig.~\ref{fig2} in which we see that in general there is a good qualitative
agreement  which becomes
also quantitatively good  at higher values of soliton backgrounds. The deviation
observed at lower values of $\rho_0$ can  be ascribed, probably, to a large
soliton width  comparable with
the size of the whole condensate. In this situation the conditions of validity of
the theory are not satisfied well enough.

\subsection{Attractive quintic interaction ($h<0$)}

For $h<0$ the integrand function in Eq.~(\ref{10-1}) is real for
\begin{equation}\label{13-3}
    \rho_0<\frac{g}{2|h|},
\end{equation}
and this inequality coincides with the condition of modulation stability of BEC,
namely the condition that the sound velocity $c_s=\sqrt{\rho_0f'(\rho_0)}$ is
real. It is worth noticing that this condition is more restrictive than the
condition $\rho_0<g/|h|$ that the nonlinearity function is positive,
$f(\rho)=g\rho-|h|\rho^2>0$. Obviously, we can discuss the soliton oscillations
for BECs satisfying the condition (\ref{13-3}) only.

In this case $Q(\rho)$ has the double zero $\rho_0$ and the simple zeroes
$\rho_+>\rho_0$ and $0<\rho_-<\rho_0$ where
\begin{equation}\label{sol4}
    \rho_{\pm}=\pm\sqrt{\left(\frac{3g}{4|h|}-\rho_0\right)^2-\frac{3v^2}{2|h|}}
    +\left(\frac{3g}{4|h|}-\rho_0\right).
\end{equation}
Hence there exist both dark and bright soliton solutions.
We shall consider here motion only of the dark soliton with the
density profile
\begin{equation}\label{sol5}
    \rho(\xi)=\rho_0-\frac{(\rho_0-\rho_-)(\rho_+-\rho_0)}
    {(\rho_+-\rho_-)\cosh^2(\kappa\xi)-(\rho_0-\rho_-)},
\end{equation}
where
\begin{equation}\label{sol6}
    \kappa=\sqrt{2|h|(\rho_0-\rho_-)(\rho_+-\rho_0)/3}.
\end{equation}
Now calculation of the number of particles $N$ and the soliton mass $m_s$ gives
\begin{equation}\label{13-4}
    N=-\sqrt{\frac6{|h|}}\arcsin\sqrt{\frac{2|h|\rho_0}{3g-4|h|\rho_0}}
\end{equation}
and
\begin{equation}\label{13-5}
    m_s=-\sqrt{\frac6{|h|}}\left[\arcsin\sqrt{\frac{2|h|\rho_0}{3g-4|h|\rho_0}}
    +\frac{2\sqrt{6|h|\rho_0(g-2|h|\rho_0)}}{3g-4|h|\rho_0}\right].
\end{equation}
Hence the effective mass of the soliton motion is equal to
\begin{equation}\label{13-6}
    m_*=\frac{m_s}{N}=1+\frac{\sqrt{6|h|\rho_0(g-2|h|\rho_0)}}
    {(3g-4|h|\rho_0)\arcsin\sqrt{\frac{2|h|\rho_0}{3g-4|h|\rho_0}}}.
\end{equation}
The condition (\ref{13-3}) means that in these expressions the argument of the
$\arcsin$ function is less than unity. (Of course, expressions (\ref{13-4})--(\ref{13-6})
can be obtained from (\ref{12-4})--(\ref{12-6}) by means of their analytical continuation
to negative values of $h$.)

For $|h|\rho_0\ll g$ we again reproduce (\ref{13-1}), as it should be.
For $g-2|h|\rho_0\ll g$ we find the limiting behavior
\begin{equation}\label{14-1}
    m_*=1+\frac2{\pi\sqrt{g}}\sqrt{g-2|h|\rho_0},\quad g-2|h|\rho_0\ll g.
\end{equation}
Thus, $m_*$ as a function of $\rho_0$ has a branching point singularity at
$\rho_{0c}=g/(2|h|)$ where the system becomes modulationally unstable.

The plot of the ratio of the trap frequency to the soliton's oscillations frequency
is illustrated in Fig.~2 (see right panel) and it demonstrates a very good agreement with
the results of the numerical simulations.

\section{Conclusion}

In this paper we have considered the dynamics of a dark soliton inside a BEC
described by the quasi-one dimensional GP equation with a parabolic trapping potential.
In particular, we have
developed the theory which permits one to study the motion of a dark soliton through
the stationary background state of BEC and to derive the formulae for the frequency of
the oscillation of deep solitons which are very effective for arbitrary form
of the nonlinearity function $f(\rho)$. When the form of $f(\rho)$ is appropriately
specified,
the formulae for the frequency of oscillations reproduce the results of all
previously studied special cases. We applied our theory to the study of the practically
interesting case of cubic-quintic nonlinearity function
$f(\rho)=g\rho+h\rho^2$ and we have demonstrated, as a new effect,  the dependence of the
frequency of the oscillations on the background density of BEC. In principle this
effect can be used for the experimental study of higher nonlinearity on
BEC's dynamics.

\subsection*{Acknowledgments}

AMK thanks the Department of Physics ``E.R.
Caianiello'' of the University of Salerno, where part of this work
was done, for the hospitality received and for financial support.
He was also partially supported by RFFI (grant 09-02-00499). MS acknowledges
partial support from a MIUR-PRIN initiative.

\end{document}